\documentstyle[prl,aps]{revtex}

\begin{document}
\draft

\title{Exciton condensates in semiconductor quantum wells emit coherent light}

\author{J. Fern\'andez-Rossier and C. Tejedor}
\address{Departamento de F\'{\i}sica Te\'orica de la Materia Condensada.
Universidad Aut\'onoma de Madrid. Cantoblanco, 28049 Madrid. Spain}

\author{R. Merlin} 
\address{Department of Physics. University of Michigan. Ann
Arbor, MI 48109-1120}

\date{\today}
\maketitle

\begin{abstract}

We show that a quasi-two dimensional condensate of optically active excitons
emits coherent light even in the absence of population inversion. This allows 
an unambiguous and clear experimental detection of the condensed phase.  We
prove that, due to the exciton-photon coupling, quantum and thermal
fluctuations do not destroy condensation at finite temperature. Suitable
conditions to achieve condensation are temperatures of a few K for typical
exciton densities, and  the use of a pulsed, and preferably circularly
polarized, laser.

\end{abstract}



A system of bosons or  paired fermions is said to be collectively condensed
when it exhibits correlations that reflect the macroscopic occupation of a
single microscopic state. Condensation can be described in terms  of a non-zero
order  parameter corresponding to the thermal average of an operator that 
creates either a boson  or a pair of fermions \cite{BEC,Forster,PW}. In the
condensed phase, the system is, locally, in a coherent superposition of
eigenstates with a  varying number of particles.\cite{PW} Condensation has been
clearly established  for superconductors and superfluid $^3$He in the case of
pairs of fermions, and for superfluid $^4$He and atomic Bose condensation in
the case of bosons.

The condensation of  an exciton gas in semiconductors has been extensively
discussed in the literature.
\cite{BEC,Kel,Ger,Nozi,Srink,Fuku,Butov,Snoke,Zhu,Flatte,Snoke2,ShamSSC,Ima,PRL1,Conti,Lai}
The transition takes place when
the system meets the so-called quantum-degeneracy criterion $n\lambda^D >1$.
Here, $\lambda= \sqrt{2\pi \hbar ^2/Mk_BT}$ is the De Broglie thermal
wavelength, $n$ is the particle density, $D$ is the number of spatial
dimensions, and $M$ is the particle mass \cite{BEC}. Thus, the criterion is met
for small $M$ at low $T$ and high $n$. Apart from $n\lambda^D$, another
relevant dimensionless quantity for excitonic systems is $n a^D$ where $a$ is
the exciton radius. For $na^D <<1$ excitons behave like weakly interacting
bosons, otherwise they must be treated as pairs of
fermions.\cite{Kel,Nozi,Srink} For $n\lambda^D >1$, condensation in the exciton
gas can be expected in two flavors depending on the exciton size: a genuine
Bose condensation for $n a^D <<1$ and, in the oppoiste limit, a BCS-like
transition as for superconductors or $^3$He. The finite lifetime of excitons
represents an obvious drawback to achieve condensation. However, we note that
typical lifetimes are long enough for  the excitons to reach a quasi
thermodynamic equilibrium, which can be  studied using time-resolved
techniques.\cite{Shah}

The condensation of optically active (OA) excitons has received considerable
experimental \cite{Fuku,Butov} and theoretical
\cite{Kel,Ger,Nozi,Srink,Fuku,Zhu,Flatte,ShamSSC,Ima,PRL1,Lai} attention in the
past several years. In such a system a crucial question arises as to the
character of the light emitted by the condensate.  Superradiance and coherent
emission have been considered in the
past\cite{Fuku,Zhu,ShamSSC,Ima,Lai} but, to the best of our knowledge, a
definitive discussion of emission properties has not been given before \cite
{Snoke2}.

In this letter, we consider the light emitted from a condensate of OA excitons,
created using a fast laser pulse, in semiconductor Quantum Well (QW)
structures. Irrespective of the process that drives the excitons to condense
after the laser is turned off, we prove that: (i) the emitted light is a
coherent electromagnetic field (EMF) of the Glauber type \cite{Glauber}; (ii) 
condensation of OA excitons can take place at finite temperatures in an
infinite and flat two-dimensional (2D) system. This is significantly different
from the case of an {\it isolated} 2D system for which Hohemberg theorem
\cite{Hohem} forbids condensation at finite temperatures. The importance of (i)
is twofold. First, the coherent emission could be used to probe exciton
condensation . Second, the emission from the condensate is expected to occur at
any density even if there is no population inversion \cite{Ima}. Therefore, it
constitutes a  physical phenomenon different from that of lasing in high
density exciton gases where coherent emission follows from population 
inversion.\cite{Flatte}

We are interested in a semiconductor QW excited by a laser pulse
with a central frequency that is resonant with the lowest absorption exciton
peak. The laser excitation leads to two related effects in that it creates an
exciton gas in the QW, and it also induces a time dependent electric dipole
moment ${\bf P}$. The latter results from the fact that the resonantly excited
QW is in a many-exciton coherent quantum state.\cite{Koch} After the laser
pulse has passed, the excited QW system will require a certain time to attain
equilibrium. Let us discuss first the normal case in which there is no
condensation.  Here, we need to consider two relaxation processes. The faster
one is the decay of the laser induced $\langle {\bf P} \rangle $ due to exciton
thermalization: the exciton coherent state is replaced by a thermal
distribution.  This is strongly supported by experiments indicating that, a
short time after the transit of the laser pulse, the exciton gas can be
described in terms of an equilibrium distribution \cite{Srink,Shah,Koch}. In
this context it makes sense to characterize the exciton gas  by a temperature
$T_X$, and by a density $n$, that is proportional to the electric dipole
fluctuation $\langle |{\bf P}|^2 \rangle$. The slower relaxation process is the
exciton recombination producing the decay of $n$. This kind of relaxation is
slow enough so that the excitons  can be considered to be, at any time, in
thermal equilibrium. Exciton recombination  gives rise to incoherent light
emission or luminescence, coming  from thermal and quantum fluctuations of
${\bf P}$ around its zero mean value. A completely different scenario occurs
when excitons condense because, as shown below, the electric dipole moment is
non zero even in the absence of an external EMF. In this case, photon emission
has a coherent component originating  in the oscillating electric dipole
moment. Since the emission from the condensate ${\bf P}$,  can be mistaken for
that of the external laser it is desirable to perform the condensation
experiments with a pulsed source.

In the following, we show that a condensate of 2D excitons  gives rise to
coherent light emission. We focus on 2D systems (as a GaAs QW) in order to
avoid a discussion of polariton effects,  that are important in bulk
semiconductors.\cite{Koch} However, an oscillating electric dipole should
also appear in a three dimensional condensate, resulting in coherent
emission for samples of dimensions comparable to the light wavelength.

The Hamiltonian of the photons plus the QW
excitons is $H=H_X+H_L+H_{XL}$ where
\begin{eqnarray}
H_X& = & T_e+T_h+V_{ee}+V_{hh}+V_{eh} \\ 
\label{excitons}
H_L& = & \sum_{{\bf q},m} \hbar cq b^{\dagger}_{{\bf q},m}b_{{\bf q},m} , \\
\label{light}
H_{XL}& = & g D_{cv} \sum_{{\bf k},{\bf q},m}
c^\dagger _{{\bf k},\sigma _e} d^\dagger _{{\bf k},\sigma _h}
b_{{\bf q},m} + h.c. 
\label{xl}
\end{eqnarray}
The exciton  component of the Hamiltonian, $H_X$,  contains all the kinetic
($T$'s) and Coulomb interaction ($V$'s) terms corresponding to electrons and
holes created by the operators $c^\dagger _{\bf k, \sigma_e}$ and $d^\dagger
_{\bf k,\sigma_h} $ where $\bf k$ is the fermion 2D wavevector and the $\sigma$'s
are the third components of the fermion angular momenta. $b^{\dagger}_{\bf
q,m}$ are the photon operators with polarization $m=\sigma_e +\sigma_h$ (OA
excitons) and momentum $\bf q$ along the direction perpendicular to the 
QW due to conservation of in-plane momentum in Eq. (\ref{xl}). 
$H_{XL}$ is the exciton-light dipolar coupling given in the rotating wave
approximation. $H_{XL}$   gives processes in which one exciton is created and
one photon is destroyed or vice versa; the total number of excitons {\em plus}
photons  is conserved.  $g$ is the light-matter coupling constant \cite{Koch}
and $D_{cv}$ is the matrix element of the dipole moment operator between the
valence and the conduction band. By definition, $D_{cv}\neq 0$ in an OA
semiconductor. We also impose that $D_{cv}$ is real and we ignore its
dependence on $\bf k$.\cite{Koch}  In GaAs QW structures, we need to consider
two classes of OA excitons characterized by the third component of its total
angular momentum $m$. Pumping by circularly polarized (left or right) light
selects $m=\pm1$. Using a mean-field approach  which neglects 
electron-hole exchange,
it has been shown \cite{Nozi,PRL1} that excitons with $m=1$ and $m=-1$ do not
couple to each other. This allows us to consider two independent coexisting
exciton gases each one coupled to a different light polarization state.

The state of exciton condensation is
defined as the emergence, without any external EMF, of  the complex order
parameter: \cite{BEC,Forster,PW,Kel,Nozi,Srink,Zhu}:
\begin{eqnarray}
\langle c^{\dagger}_{\bf k, \sigma_e} d^{\dagger}_{\bf
k,\sigma_h} \rangle \equiv f_{\bf k,m} e^{-i \phi_{m}} \label{sgsb2}
\end{eqnarray}
where $\langle \rangle$ stands for the statistical average. An essential
feature of the condensation is that the phase $\phi_m$ of the order parameter
is independent of $\bf k$, i.e., the phase is the same  across the exciton gas,
a signature of long range order. The  magnitude of the order parameter $f_{\bf
k,m}$ can be calculated in two different situations. In the mean field BCS
approximation, valid for  arbitrary densities, $f_{\bf k,m}= u^{m}_{\bf
k}v^{m}_{\bf k}$ where $u^{m}_{\bf k}$ and $v^{m}_{\bf k}$ are the coefficients
of a BCS wave function to be calculated for each particular
problem.\cite{Zhu,PRL1} In the limit of low densities, excitons behave like
nearly free bosons and, thus,  $f_{\bf k,m}=\sqrt{N_m}$, where $N_m$ is the
number of excitons of type $m$ \cite{BEC,Forster,PW}. Our results do not depend
on the approximations made to calculate $f_{\bf k,m}$. 

Let us now discuss the behavior of the electric dipole moment in terms of 
its circular components  ${\bf P}_m= {\bf P}_x+im{\bf P}_y$ with $m=\pm 1$ 
which couple to photons. They are given by \cite{Koch}
\begin{eqnarray} 
P_{m} \equiv
\sum_{\bf k,m} D_{cv} c^\dagger _{\bf k, \sigma_e} d^\dagger _{\bf k, \sigma_h}
+ h.c. \label{pol} 
\end{eqnarray}
Combining Eqs. (\ref{sgsb2}) and (\ref{pol}), we get that, for an $m$ type
exciton condensation, the mean value of $P_m$ is non zero: 
\begin{eqnarray}
\langle P_m \rangle= F(N_m) cos(\phi_m)
\end{eqnarray}
where $F(N_m)=2 D_{cv} \sum_{\bf k} f_{\bf k,m}$. It follows that exciton
condensation implies that $\langle P \rangle$ is non zero with no external
EMF. In the presence of an external EMF, the response function of the
condensate  exhibits a singular behavior.\cite{Ger} In the normal case, the
many exciton quantum states are assumed to be eigenstates of the exciton number
operator and, therefore, both the condensation order parameter and $\langle P
\rangle$ are equal to zero.

We now show that the condensation-induced ${\bf P}$ oscillates with a
frequency identical to that of the lowest exciton state This oscillating
dipole emits coherent light. A full quantum mechanical calculation of the
time evolution of ${\bf P}$ is given in reference \cite{Koch}. The same
result can be obtained in a simpler way from the equation of motion of the
phase given by \cite{PW}
\begin{eqnarray} 
\frac{d}{dt} \phi_m = \hbar ^{-1}\frac{\partial}
{\partial N_m} E(N_m, \phi _m)
\label{eqnmotion}
\end{eqnarray}
where $E(N_m, \phi _m)=\langle H_X + H_{XL} \rangle$. There are three
contributions to $E(N_m, \phi _m)$. From large  to small, we have the single
exciton contribution  $(E_g - E_b)N_m$ where $E_g$ is the semiconductor gap and
$E_b$ is the binding energy, the exciton-exciton interaction and the
exciton-light contribution $\langle H_{XL} \rangle$.  In a GaAs QW, the first
one is approximately 1.5 ev giving rise to an  electric dipole oscillation
period in the scale of femtoseconds.  The many-exciton corrections simply give
a renormalization  $\overline{E}_b-E_b$ of the exciton binding energy of the
order of 1meV.\cite{Srink,Zhu,PRL1}

Finally, the exciton-light interaction which gives a contribution proportional 
to $cos(\phi _m)$; this term is negligible compared with the single  exciton
term because $g$ is typically three orders of magnitude smaller than the gap. 
In contrast to the phase dynamics, $f(N_m)$ is a very slow quantity because
the density varies in  a ps scale.\cite{Shah} Hence, we have a time
dependent electric dipole moment
\begin{eqnarray} 
\langle P_m(t)\rangle
= F(N_m(t)) cos \left( \frac{(E_g - \overline{E}_b) t}{\hbar} + \phi_m(0))\right)
\end{eqnarray}
which leads to an oscillating EMF. Our next
task is to provide a quantum description of the emission process and show  that
the photon field emitted by the condensate is in a quantum mechanical coherent
state.\cite{Glauber} We only consider the emission resulting  from the 
condensation-induced electric dipole, neglecting  the emission  due to
fluctuations. In order to do that, we just need the equation of motion of the
photon operators, replacing the electric dipole operator by  its mean values.
The light part of the Hamiltonian becomes
\begin{eqnarray}
H_L+H_{XL}=
\sum_{\bf q,m} \left[ \hbar cq  b^{\dagger}_{\bf q,m}b_{\bf q,m} +
\frac{gF(N_m(t))}{2} \left(e^{-i \phi_{m}(t)} b_{\bf q,m} + h.c.  \right)
\right]. 
\label{AO}
\end{eqnarray}
This term  corresponds to a  set of
non interacting forced harmonic oscillators which can be solved for any
$F(N_m(t))e^{-i\phi_{m}(t)}$. Assuming that the condensation starts  at $t=0$,
{\em i.e.} that the forcing term is zero for $t<0$, the exact single mode
photonic field for $t>0$ is \cite{Galindo}: 
\begin{eqnarray} |\Xi_{\bf
q,m}(t) \rangle & = & e^{\Theta_{q,m}(t)-icq n_{\bf q,m} t} e^{i K_{q,m}(t) 
b^{\dagger}_{\bf q,m}} e^{iK_{q,m}^{*}(t)b_{\bf q,m}} |\Xi_{\bf q,m}(0) \rangle
\nonumber \\ K_{q,m}(t) & = & \frac{g}{2\hbar} \int_{0}^{t} F(N_m(s))
e^{i[cqs -\phi_m(s)]} ds , \nonumber \\ \Theta_{q,m}(t) & = & \int_{0}^{t}
K_{q,m}(s) \frac{dK_{q,m}^{*}(s)}{ds}ds,
\end{eqnarray}
where $n_{\bf q,m}\equiv b^{\dagger}_{\bf q,m} b_{\bf q,m}$. Assuming that 
there are no
photons before the exciton condensation, the photon field for $t>0$ is
\begin{eqnarray}
|\Xi(t) \rangle =e^{-i H_L t/\hbar} \prod_{\bf q,m}
e^{\Theta_{q,m}(t)} e^{i K_{q,m}(t) b^{\dagger}_{\bf q,m}} |0 \rangle
\label{coherent} 
\end{eqnarray} 
which represents a Glauber coherent state \cite{Glauber}. For coherent states,
the expectation value of the electric field is equal to the classical
electric field generated by an oscillating ${\bf P}_m$. All this is a
consequence of 
\begin{eqnarray}
\langle \Xi(t)|b_{\bf q,m}|\Xi(t) \rangle \neq 0 \label{bneq}
\end{eqnarray}
which holds even if the photon field is not in its ground state at $t=0$.
This proves  our first claim that there is a coherent emission coming from  an
exciton condensate. Obviously, the state varies with time for the emission of
one photon  removes one exciton from the condensate. Also, note that Eq.
(\ref{bneq}) is not verified in the case of luminescence.

The connection between condensation and $\langle P \rangle \neq 0$ was
emphasized by Sham and collaborators for an excitonic insulator.\cite{Shamfk}
As  the excitonic insulator is in a true equilibrium state, the electric dipole
moment associated  with the condensation is time independent, in contrast to
our case. A non zero $\langle P \rangle$ in an OA exciton condensate  is also
mentioned  in reference \cite{ShamSSC}, although these  authors did not discuss
the issue of the coherence of  the emitted EMF . 

We  now  consider the possible existence of 2D condensates at finite
temperatures. Hohemberg rigorously proved that, under certain conditions,
condensation  is impossible  in 2D systems at finite temperatures
\cite{Hohem}. His proof relies on the divergence of the thermal depletion
of the order parameter $\langle c^{\dagger}_{\bf k} d^{\dagger}_{\bf
k}\rangle $ which is bigger than the integral \cite{Forster,Hohem}:
\begin{eqnarray} 
\int \frac{d^2{\bf
k}}{\langle[[\rho_{\bf k} ,H_X],\rho^{\dagger}_{\bf k}]\rangle} \label{in}
\end{eqnarray} 
where $\rho_{\bf k}$ is the $\bf k$ Fourier component of the density operator.
Since $H_X$ conmutes with the number operator,  it is easy to demonstrate
\cite{Pines} the f-sum rule $\langle[[\rho_{\bf k} ,H_X],\rho^{\dagger}_{\bf
k}]\rangle= \hbar ^2 k^2 (N_{0,1}+N_{0,-1})/2M$. If $H_X$ is the only term in
the Hamiltonian, the integral (\ref{in}) contains an infrared singularity
and the thermal depletion of the order parameter must be infinite, i.e.,
there is no condensation at finite temperature in 2D. For OA excitons,  
there is an additional  contribution $H_{XL}$ which must be included in the
analysis. This  gives rise to an extra term in the denominator of the
integral (\ref{in}) which suppresses the singularity. For a given value
of $m$, the new term takes the form
\begin{eqnarray}
\langle[[\rho_{\bf
k} ,H_{XL}],\rho^{\dagger}_{\bf k}]\rangle  = gD_{cv} \sum_{{\bf q},{\bf
p}} \langle b^\dagger _{\bf q} \rangle \left( \langle c^\dagger _{\bf
p+k}d^\dagger _{\bf p-k} \rangle +\langle c^\dagger _{\bf p-k}d^\dagger
_{\bf p+k}  \rangle -2\langle c^\dagger _{\bf p}d^\dagger _{\bf p} \rangle
\right) + h.c.  \label{sqex}
\end{eqnarray}
The infrared divergency in the integral (\ref{in}) disappears because Eq.
(\ref{sqex}) depends linearly on $k$.\cite{Kel} This term is non zero due to
Eq. (\ref{sgsb2}) which implies Eq. (\ref{bneq}) as shown above. In other
words, long wavelength thermal fluctuations do not destroy the long range order
because the condensed excitons can recombine emitting coherent light. The
infrared divergency is associated  with restoration of the  gauge symmetry 
by the gapless Goldstone modes of infinite wavelength. The coupling between
excitons and photons breaks this symmetry  and therefore there is no
gapless Goldstone modes and no divergency. The conclusion is: {\em optically
active exciton condensation is  possible in 2D at finite temperatures}

While we have established that a condensate of 2D OA excitons can exist at
finite temperatures, the prediction of the critical temperature $T_c$ is a
difficult matter. Here, we give upper bounds based on our physical
understanding of the problem. There are two relevant energy scales which
are upper bounds for $T_c$\cite{BEC,Conti}: the quantum degeneracy
temperature, $T_{QD}= 2\pi \hbar^2 n/M$ below  which the thermal wavelength
$\lambda_{B}$ is larger than the interparticle distance and the pairing
energy, $\Delta$, of the condensate. Numerical calculations in the BCS
approximation \cite{Zhu} show that, for densities below $10^{11} cm^{-2}$
and $d<100 \AA$,  $\Delta > T_{QD}$. Hence, exciton condensation
experiments should be performed  at temperatures below $T_{QD}$. For
exciton  masses of GaAs QW, $T_c$ (in K) $\simeq 10^{-10} \times n $(in
cm$^{-2}$) so that a typical $T_c$ is a few degrees K.

A plausible alternative to  increase the critical temperature for condensation 
is to use unbalanced populations $n_{+1} \neq n_{-1}$ of $\pm 1$ excitons . To
enhance the quantum degeneracy and, consequently, $T_{QD}(m) \propto N_m$ it is
convenient to have all the excitons in the same state  $m$.  and, for this,
one can use circularly polarized  light .\cite{Shah}  Typically,
populations tend to balance in 50 ps for a symmetric QW. This
relaxation time can be increased in an  asymmetric QW because the
relaxation mechanism is proportional to the  electron-hole overlap.
Moreover, in this case a fully polarized ground state is expected
\cite{PRL1} increasing the quantum degeneracy. In some cases, both $+1$ and
$-1$ excitons can be condensed\cite{PRL1} and we expect that new physical
phenomena related to the coherence between the $+1$ and the $-1$ phases 
will be uncovered \cite{gottin}. However, it  should be noted that the
critical temperature for this two-component condensation is still lower
than the one associated to the single-component  condensation. 

In summary, a condensate of OA excitons in a 2D heterostructure will emit
coherent light even in the absence of population inversion. This effect can be
used to confirm unambiguously exciton condensation. The coherent emission is an
essential ingredient of condensation because it cancels long wavelength
fluctuations which  prevent condensation in other 2D systems.

The authors would like to thank L. Sham for a critical reading of the
manuscript and D. W. Snoke for  helpful comments. Work supported in part by
MEC of Spain under Contract Number PB96-0085, by the Fundaci\'on Ram\'on
Areces and by the U. S. Army Research Office under Contract Number
DAAH04-96-1-0183.  During the early stages of this work, R.M. was a Visiting
Iberdrola Professor at the Universidad Aut\'onoma de Madrid.


\begin{references}

\bibitem{BEC} {\it Bose-Einstein Condensation}, edited by A.
Griffin, D. W. Snoke, and S. Stringari (Cambridge, Cambridge, 1995).

\bibitem{Forster} D. Forster, {\it Hydrodynamic Fluctuations, Broken Symmetry,
and Correlation Functions} (Benjamin, New York, 1975).

\bibitem{PW} P. W. Anderson, {\it Basic Notions of Condensed Matter
Physics} ( Benjamin-Cummings, Menlo Park, 1984).

\bibitem{Kel} L. V. Keldysh and Yu. V Kopaev, Fiz.\ Tverd.\ Tela {\bf 6}, 2791
(1964) [Sov.\ Phys.\ Solid State {\bf 6}, 2219 (1965)]; D. Jerome, T.  M. Rice,
and W. Kohn, Phys.\ Rev.\ {\bf 158}; E. Hanamura and H. Haug, Phys. Rep., {\bf
33}, 209 (1977). 

\bibitem{Ger}V. A. Gergel {\em et al.}, Zh. Eksp. Teor. Fiz., {\bf 53}, 544
(1967) [Sov.\ Phys.\ Jetp {\bf 26}, 354 (1968)]. 

\bibitem{Nozi} C. Comte and P. Nozi\`{e}res, J. Physique {\bf 43}, 1069
(1982); P. Nozi\`{e}res and C. Comte,{\em ibid} {\bf 43}, 1083 (1982). 

\bibitem{Srink} H. Haug and S. Schmitt-Rink, Prog. Quant. Electr.,  {\bf 9}, 3
(1984).

\bibitem{Fuku} T. Fukuzawa {\em   et al.}, Phys. Rev.
Lett.{\bf 64}, 3066 (1990); Surface Sc. {\bf 228}, 482 (1990); Kash {\em et
al.}, Phys. Rev. Lett
{\bf 66}, 2247 (1991)

\bibitem{Butov} L.V. Butov {\em et al.}, Phys. Rev. Lett. {\bf 73}, 304
(1994). 

\bibitem{Snoke}  J.P Wolfe {\em et al.} in  \cite{BEC} 

\bibitem{Zhu} X. Zhu {\em et al.}, Phys. Rev. Lett. {\bf 74}, 1633 (1995);
P. B. Littlewood and X. Zhu, Physica Scripta {\bf T68}, 56 (1996). 

\bibitem{Flatte} M.E. Flatte, {\em et al.}, Appl. Phys. Lett., {\bf 66},
1313 (1995). 

\bibitem{Snoke2} See, e.g., D. W. Snoke, Science {\bf 273}, 1351 (1996), and
references therein; S. A. Moskalenko and D. W. Snoke, {\it Bose-Einstein
Condensation of Excitons and Biexcitons} (Cambridge University Press), to be
published.

\bibitem{ShamSSC} Th. Ostreich {\em et al.}, Solid St. Commun., {\bf 100}, 325
(1996).

\bibitem{Ima} A. Imamoglu {\em et al.} Phys. Rev. A{\bf53}, 4250(1996) 

\bibitem{PRL1} J. Fern\'andez-Rossier and C. Tejedor, Phys. Rev. Lett. {\bf
78}, 4809 (1997).

\bibitem{Conti} S. Conti, {\em et al.}, Phys. Rev. B{\bf 57}, R6864 (1998)

\bibitem{Lai} B. Laikhtman, Europhys. Lett., to be published. 

\bibitem{Shah} J. Shah, {\em Hot Carriers in Semiconductor Nanostructures},
(Academic Press, San Diego, 1992). 

\bibitem{Glauber} R.J. Glauber, Phys. Rev.  {\bf 131}, 2766 (1963)

\bibitem{Hohem}  P.C. Hohemberg, Phys. Rev. {\bf 158}, 383 (1967).

\bibitem{Koch} H. Haug and S.W. Koch, {\it Quantum Theory of Optical and
Electronic Properties of Semiconductors} (World Scientific, London, 1993).

\bibitem{Galindo} A. Galindo and P. Pascual,  {\em Quantum Mechanics}, (Springer
Verlag, Berlin, 1991). 

\bibitem{Shamfk} T. Portengen {\em   et al.}, Phys. Rev. Lett. {\bf 76} (1996);
Phys. Rev. B{\bf 54}, 17452 (1996).

\bibitem{Pines} D. Pines and P. Nozieres, {\it The theory of Quantum Liquids,
vol I} (Addison-Wesley, Reading, 1966). 

\bibitem{gottin} J. Fern\'andez-Rossier and C. Tejedor, Phys. Stat. Sol. (a), 
{\bf 164}, 343 (1997). 
\end{references}
\end{document}